\documentclass[]{spie}  

 \usepackage{gensymb}
\usepackage{amsmath,amsfonts,amssymb}
\usepackage{graphicx}
\usepackage[colorlinks=true, allcolors=blue]{hyperref}
\usepackage{textcomp}
\usepackage{xcolor}

\title{DiskFM: A Forward Modeling Tool for Disk Analysis with Coronagraphic Instruments
}

\author[a]{Johan Mazoyer}
\author[b]{Pauline Arriaga}
\author[c]{Justin Hom}
\author[d]{Maxwell A. Millar-Blanchaer}
\author[e,f]{Christine Chen}
\author[d]{Jason Wang}
\author[g,h]{Gaspard Duchêne}
\author[c]{Jennifer Patience}
\author[e]{Laurent Pueyo}

\affil[a]{LESIA, Observatoire de Paris, Université PSL, CNRS, Sorbonne Université, Université de Paris, 92195 Meudon, France}
\affil[b]{Department of Physics \& Astronomy, University of California, Los Angeles, CA 90095, USA}
\affil[c]{School of Earth and Space Exploration, Arizona State University, P.O. Box 871404, Tempe, AZ 85287, USA}
\affil[d]{Department of Astronomy, California Institute of Technology, Pasadena, CA 91125, USA}
\affil[e]{Space Telescope Science Institute, 3700 San Martin Drive, Baltimore, MD 21218, USA}
\affil[f]{Department of Physics and Astronomy, The Johns Hopkins University, Baltimore, MD 21218, USA}
\affil[g]{Astronomy Department, University of California, Berkeley, CA 94720, USA}
\affil[h]{Univ. Grenoble Alpes, CNRS, IPAG, 38000 Grenoble, France
}

\authorinfo{Further author information: johan.mazoyer@obspm.fr}

\begin{document} 
\maketitle

\begin{abstract}
Because of bright starlight leakage in coronagraphic raw images, faint astrophysical objects such as exoplanets can only be detected using powerful point spread function (PSF) subtraction algorithms.  However, these algorithms have strong effects on faint objects of interest, and often prevent precise spectroscopic analysis and scattering property measurements of circumstellar disks. For this reason, PSF-subtraction effects is currently the main limitations to the precise characterization of exoplanetary dust with scattered-light imaging. 

Forward modeling techniques have long been developed for point source objects (Pueyo 2016). However, Forward Modeling with disks is complicated by the fact that the disk cannot be simplified using a simple point source convolved by the PSF as the astrophysical model; all hypothetical disk morphologies must be explored to understand the subtle and non-linear effects of the PSF subtraction algorithm on the shape and local geometry of these systems. Because of their complex geometries, the forward-modeling process has to be repeated tens or hundred of thousands of times on disks with slightly different physical properties. All of these geometries are then compared to the PSF-subtracted image of the data, within an MCMC or a Chi-square wrapper. 

In this paper, we present here {\tt DiskFM}, a new open-source algorithm included in the PSF subtraction algorithms package {\tt pyKLIP}. This code allows to produce fast forward-modeling for a variety of observation strategies (ADI, SDI, ADI+SDI, RDI). {\tt pyKLIP} has already been used for SPHERE/IRDIS and GPI data. It is readily available on all instruments supported by pyKLIP (SPHERE/IFS, SCExAO/CHARIS), and can be quickly adapted for other coronagraphic instruments.
\end{abstract}


\section{INTRODUCTION}
\label{sec:intro}  

The main challenge of direct imaging of exoplanets and disks is the large flux ratios between faint circumstellar objects and their host stars. Recent progress has been achieved with the combination of extreme-adaptive optics systems with state-of-the-art coronagraphs in the latest generation of ground-based instruments: VLT/SPHERE \cite{Beuzit2019}, Gemini/GPI \cite{Macintosh2014} and Subaru/SCExAO \cite{Jovanovic2015, lozi_scexao_2018}. Great improvements to the final performance in contrast have also been achieved by the development of observation strategies and of a posteriori image treatment algorithms to subtract the point spread function (PSF) of the star in coronagraphic images. 

These algorithms are based on the assembly of a library of PSF references. For each science frames, a linear combination of the PSFs in the library can then extracted to fit and subtract the stellar speckles. Several techniques have been designed to assemble a PSF library. In some cases, the collection of reference images can sometimes be obtained using observations of calibration stars with no known circumstellar companion (Reference Differential Imaging; hereafter RDI). This technique is favored for very stable PSFs (e.g. from space HST/STIS or HST/NICMOS) because the reference calibration star must be observed at a different time. However, for Ground-based telescopes in Near-IR, rapid atmospheric turbulence often makes this observation strategy risky and less effective (with the possible exception of very large reference PSF libraries). Observation strategies have then been developed using a \textit{a priori} known difference during the observation between the faint astrophysical objects that we seek to detect (disk or exoplanet) and the frame attached to the star residuals (speckles). Each image can then be alternatively treated as a science frame or included in the reference library for other frames in the sequence. Among these techniques, we can cite Angular Differential Imaging (ADI\cite{marois_angular_2006}), which uses the azimuthal motion of the astrophysical signal with respect to the speckles and Spectral Differential Imaging (SDI\cite{racine_speckle_1999}), which uses the radial motion of the astrophysical signal with respect to the speckles, for example in the case of Integral Field Spectrometers (IFS). Finally, some techniques use the inherent properties of the incoming light of the potential astrophysical objects, such as Polarisation Differential Imaging (PDI\cite{kuhn_imaging_2001}) or Coherent Differential Imaging (CDI \cite{jovanovic_review_2018}).

Once these libraries have been assembled, several algorithms were designed to create an idela PSF to fit and subtract in the science frame. From the initial algorithms classical ADI (cADI \cite{marois_angular_2006}), Locally Optimized Combination of Images (LOCI \cite{lafreniere_new_2007}) and the principal component algorithm (PCA \cite{amara_pynpoint_2012}) also called Karhunen-Loève Image Processing (KLIP \cite{soummer_detection_2012}), several optimizations have been proposed, for example adapting the size of the reduction zones \cite{marois_tloci_2014} or changing how linear components are selected \cite{gomez_gonzalez_low-rank_2016}. Other methods have been developed which use prior knowledge on the speckle noise distribution (e.g. MEDUSAE\cite{ygouf_simultaneous_2013} or PACO\cite{flasseur_exoplanet_2018}) to separate them from the planet. 

However, speckle subtraction algorithms create distortion of the astrophysical signal caused by either the aggressive subtraction of the object mistaken for speckle noise (over-subtraction) or, when the planet or disk is present in the library (ADI, SDI), due to the object subtracting itself (self-subtraction). This is a problem not only in the context of a detection, but also for the analysis of a system, because the astrometry (for a planet) or shape (for a disk) and photometry of the object can be altered. Pueyo (2016)\cite{pueyo_detection_2016} derived a second order approximation of the self- subtraction for KLIP, called forward modelling (FM). This has since been applied in the context of Match Filter \cite{ruffio_improving_2017} to consistently look for planets on large sets of data. However, all of these algorithms have mainly been applied to planet detection and characterization and not for circumstellar disks.

In this paper, we present a KLIP-FM-based algorithm developed specifically for disks, {\tt DiskFM}. This algorithm was successfully used to analyze the HR 4796 A disk for GPI and SPHERE/IRDIS data \cite{chen_multiband_2020, arriaga_multiband_2020}. This algorithm is now publicly available, included in the python based {\tt pyKLIP}\footnote{Available under open-source license at \url{https://bitbucket.org/pyKLIP/ pyklip}.} package \cite{wang_pyklip_2015}. This package has been originally designed to be multi-instrument and most features (including {\tt DiskFM}) are readily available for GPI, SPHERE, SCExAO coronagraphic instruments. 

In Section \ref{sec:purpose}, we will recall the specific challenges of PSF-subtraction techniques in the context of circumstellar disks. These specific challenges makes this kind of approach very costly in computer time and resources. We describe in Section \ref{sec:implementation} the implementation of {\tt DiskFM} and the choices that were made to optimize the disk forward-modeling. The cost in computer time and resources is probably one of the reasons than almost none of the PSF-subtraction algorithms available have tried to show their performance on a large range of injected disk geometries or reduction parameters, which is fairly common for exoplanets. In Section \ref{sec:performance}, we will test {\tt DiskFM} on 5 different simulated disks with two different observation methods (ADI/RDI) and three different reduction parameters for each.

\section{Purpose of {\tt DiskFM}: Impact of PSF-subtraction algorithms on extended structures}
\label{sec:purpose}

Circumstellar (protoplanetary and debris) disks are significantly harder to analyze in the context of PSF subtraction for two reasons: 1. their extended shape makes them extremely sensitive to the ADI/SDI self-subtraction effect, and 2. their complex structures require a large number of parameters ($\sim$ 10) to be accurately modelled. 

\begin{figure}
\begin{center}
 \includegraphics[width = \textwidth]{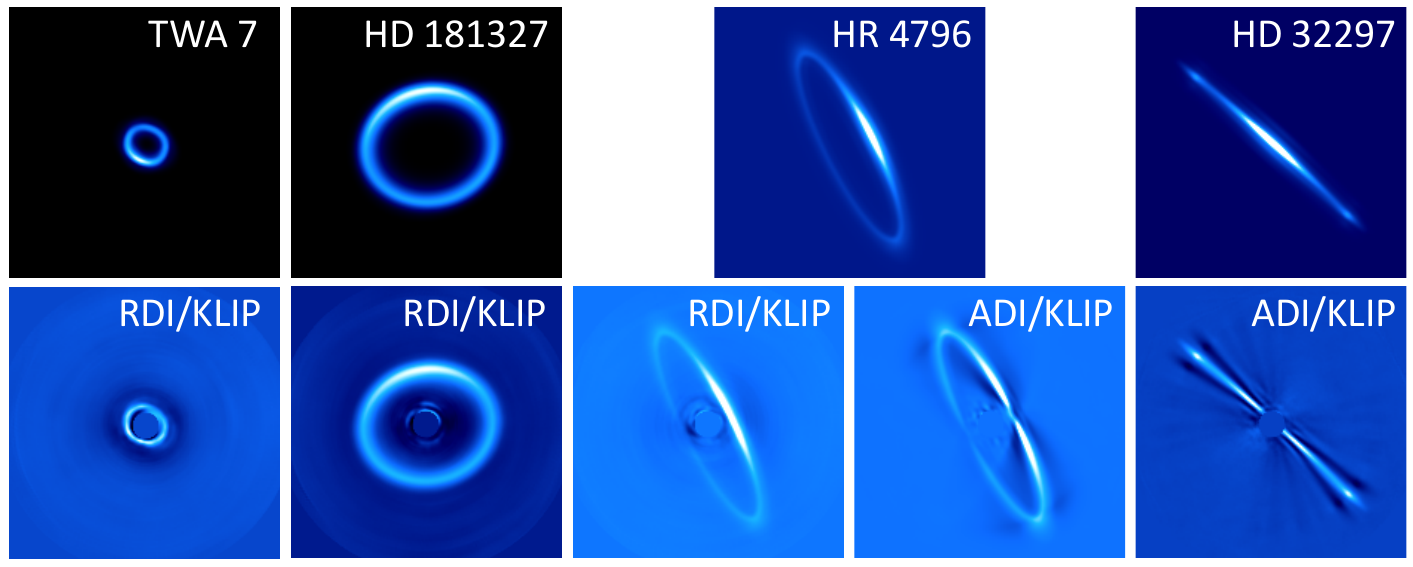}
 \end{center}
\caption[plop]
{\label{fig:show_oversub_selfsub} Impact of ADI and RDI on simulated debris disks. \textbf{Top:} Original disk (after convolution by telescope). \textbf{Bottom}: {\tt DiskFM} Forward Models in RDI and ADI modes. \textbf{Left:} KLIP/RDI, only the effect of over-subtraction is impacting the disk. This is most critical for disks with parts in zones with high speckle noise (TWA 7 and small separation regions of HR 4796 A). \textbf{Right:} KLIP/ADI, both the over-subtraction and self-subtraction effects are impacting the disk. The effect is present at all separations and for all geometries. In the case of face-on disks (TWA 7 or HD 181327), self-subtraction due to KLIP/ADI (not shown here) would often prevent detection.}
\end{figure}

Because disks are extended structures, they are very sensitive to the effect of PSF-subtraction algorithms (over- and self-subtraction). First, planets are very specifically localized in the focal plane, and often a specific set of reduction parameters can be chosen to optimize the signal to noise of the detected candidate once identified, depending on their separation to the host star and photometry. On the other hand, disks often extend at large and small separations and their photometry is locally dependent with often a ratio of a few tens from the brightest and faintest regions of the disk. Second, in the case of ADI and SDI, the astrophysical object itself is in the PSF library, which leads to self-subtraction. In the case of planet, this can be minimized, for example by using an exclusion parameter ensuring that the planet position in the Reference Library frames is distant enough from the planet in the science frames. However, when the astrophysical object is widely extended in the focal plane, exclusion parameters are not as effective and self-subtraction is very significant. Disk analysis without carefully taking into account the effect of self-subtraction has already lead to the publication of nonexistent structures (e.g. the 'streamers' of HR 4796 A \cite{thalmann_images_2011} were later found to be the effect of self-subtraction \cite{milli_impact_2012}). In the worst cases, face-on disks are totally self-subtracted and cannot be detected. This was particularly clear in the GPI disk survey analysis \cite{esposito_debris_2020} where out of 26 detected disks, all of the disks with low inclination (i < 70\textdegree) are detected only in polarized intensity (PDI reduction) and not in total intensity (ADI reduction).  Effects of over-subtraction (in RDI and ADI/SDI cases) or self-subtraction (only in ADI/SDI cases) are shown for a different disk geometries in Fig.~\ref{fig:show_oversub_selfsub}.

PSF-subtraction effects are currently one the main limitations to the precise characterization of exoplanetary dust with scattered-light imaging. Geometrical disk parameters are affected but can sometimes be extracted at a lower precision directly from the self-subtracted image. However, limitations due to self-subtraction become extremely clear when we try to analyze more subtle properties. Indeed, the absolute photometry (and therefore spectroscopy) or local variations of the photometry as a function of the observing angle cannot be reliably extracted in a self-subtracted disk image. Scattering phase functions (SPFs), linking the scattering properties of the dust to the angle of observation, have long been acknowledged as a unique tool for accessing the size, composition and shape of the grains \cite{hughes_debris_2018} in solar system or exoplanetary dust.

The first approach to remedy this problem is to use conservative reduction parameters (as opposed to aggressive in the case of planet)\cite{thalmann_imaging_2010}: large optimization regions or full frame reduction instead of locally optimized subtraction and/or the selection of only the first PCA modes (hereafter KL modes) in PCA/KLIP (3-10 compared to a few tens in the case of planets). However, this non-aggressive approach leaves a lot of speckle noise in the images, which also impacts the disk analysis. Other approaches have included masking the disk in the library to minimize self-subtraction \cite{milli_near-infrared_2017, perrin_polarimetry_2015} at the expense of Signal-to-Noise. Unfortunately, this is only possible for certain specific disk geometries and only minimizes self-subtraction rather than prevent it.

Several authors have directly analyzed the effects of over-subtraction and self-subtraction on disks depending on their geometries, as in Milli et al. (2012)\cite{milli_impact_2012} for KLIP/PCA and Esposito et al. (2014)\cite{esposito_modeling_2014} for LOCI. Finally, Pueyo (2016)\cite{pueyo_detection_2016} analytically measured the effect of KLIP on astrophysical objects for the first 3 terms of the derivation which allowed for generalized and precise forward-modeling. However, producing a FM on an extended disk can be time consuming. 

Early uses of this technique for disks as in Mazoyer et al. (2014)\cite{mazoyer_is_2014} included the determination of the structure of a disk by geometrical analysis to produce a single forward model from a fiducial model. We then analyzed the ``photometric" correction to be applied based on the ratio of the fiducial model divided by its FM. However, this approach assumed 1) that geometric and photometric properties of the disks are independent of each other, and 2)that the self-subtraction is identical for all disks of a given geometry. Both of these assumptions have proven to be wrong for a precise analysis. 

By definition, self-subtraction is the effect of the disk on itself and bright parts of the disk will have more impact than faint parts of the disk. Therefore, the photometry of the disk also impacts the FM. We conclude that we need to know the best disk model to measure the exact impact of the disk on itself. For this problem, the exploration of a large portion of the parameter space using a Chi-square or Markov chain Monte Carlo (MCMC) analysis is well suited. This is the approach of this paper. 

Another approach taken involves the selection of the modes in a Non-negative Matrix Factorization reduction (NMF\cite{gonzalez_vip_2017}). Ren et al. (2018) \cite{ren_non-negative_2018} shows that there is a specific decomposition of the PSF on the PSF library that minimizes the impact of over-fitting the disk in RDI. A recent improvement\cite{ren_using_2020} of this algorithm shows its impressive performance in ADI+SDI+RDI to recover the disk image with theoretically negligible alteration of circumstellar sources. Another interesting approach, presented in Pairet et al. (2018)\cite{pairet_reference-less_2018}, is the subtraction of the model in the library directly before doing a more classical ADI. By iterating, the algorithm slowly removes the disk in the library and minimize self-subtraction. This is currently very slow, but most likely improvable. The same team recentely published another algorithm MAYONNAISE\cite{Pairet_2020} to deconvolve the signal with promising results. 

Even assuming that we can find a method that removes all speckles noise and leaves the disk perfectly intact, we argue that we need a careful exploration of large portions of the parameter space. Indeed, geometric parameters and local photometry (i.e. SPF) are not independent. Most notably, the disk stellar offset can have an impact on the local photometry of the disk in scattered light (pericenter glow) and recent work\cite{olofsson_challenge_2020} has shown that vertical thickness of debris disks can affect the determination of their SPFs (for disks with inclinations larger than 60\textdegree, which are the majority of disks detected from the ground). Therefore, the method we use in this paper is a careful identification of the biases introduced by KLIP/ADI and KLIP/RDI using KLIP-FM formalism optimized in speed and efficiency in order to be included in an MCMC wrapper. 

\section{Implementation details}
\label{sec:implementation}

The exact analytical description of the KLIP-FM code can be found in Pueyo (2016)\cite{pueyo_detection_2016}. We used the same algorithm as for a planet, already coded in {\tt pyKLIP}. As noted in Pueyo (2016), ``in practice FMing with disks is complicated by the fact that [they] cannot be simplified using a simple PSF as the astrophysical model: every hypothetical disk morphology must be explored". Indeed, because of their complex geometries, the calculation of the FM has to be repeated many times on disks with slightly different parameters. All of these geometries are then compared to the reduced image of the data, within an MCMC or a Chi-square wrapper. Once measured for a set of reduction parameters, the Karhunen-Loève (KL) basis does not change and can be saved in a file. 
For a new disk model, the forward-modeling is therefore only an array reformatting and a matrix multiplication, which can be optimized for calculation in a few seconds. 

This is the main purpose of the {\tt DiskFM} code. The FM is calculated once and the KL basis is saved in a Hierarchical Data Format (HDF5) ``.h5" file. The user can later load this file and recalculate a FM in a few seconds only. In previous versions, only the KL coefficients were saved, not the initial data set itself (input images). This was problematic because the same KL coefficients could be used with slightly different datasets (such as the order of the frames not being identical), providing incorrect FMs. We recently updated the code, and all the information necessary is now saved inside the HDF5 file, including initial frames and reduction parameters. 
As a result, only the HDF5 file and a new model are necessary to produce a FM in the same conditions after an initial reduction is performed.

A {\tt DiskFM} tutorial can be found on the pyKLIP page\footnote{Available at \url{https://pyklip.readthedocs.io/en/latest/diskfm_gpi.html}}. {\tt DiskFM} currently only supports KLIP/ADI, KLIP/SDI, KLIP/ADI+SDI and KLIP/RDI reductions (but currently not KLIP RDI+ADI/SDI or non-KLIP reduction methods). {\tt DiskFM} has currently been used for SPHERE/IRDIS data, GPI IFS mode data\cite{chen_multiband_2020} and GPI pol mode data\cite{arriaga_multiband_2020}. {\tt pyKLIP} supports data from various instruments and {\tt DiskFM} should work without change with SPHERE/IFS and SCEXAO/CHARIS\cite{Groff_charis_2017}. A main goal has been to optimize the speed of this algorithm for ADI or RDI so that it can be incorporated in an MCMC or Chi-square wrapper. {\tt DiskFM} only provides the tools to do the forward modeling and does not include disk modeling tools or an MCMC wrapper. For the latest version, the one used by the author using {\tt emcee} package \cite{foreman-mackey_emcee:_2013} is freely available online \footnote{Available at \url{https://github.com/johanmazoyer/debrisdisk_mcmc_fit_and_plot.git}}. 

Time estimation of {\tt DiskFM} is complicated to extrapolate for every scenario because it is heavily dependent on the number of pixels in your science frame (which drives the FM matrix size) as well as the number of science frames in your sequence (and of course, computer capability of the user). One is encouraged to limit the measurement to zones where the disk has been detected using {\tt pyKLIP} inner working angle and outer angle working parameters. Binning the science frame can also be used to decrease computation time. In all of the cases, we applied (ADI and RDI reduction for extended disks in normal science sequence for SPHERE and GPI), the FM measurement has been shorter or comparable to the time necessary to produce our disk model with a simple geometric code with no physical dust grain scattering model\cite{millar-blanchaer_imaging_2016}. 
This means that the FM usually, at worse, doubles the time required to explore the parameter space.

If you input multi-wavelength dataset (e.g. IFS cubes sequences) and model, {\tt DiskFM} will produce a multi-wavelength FM. Multi-wavelength disk FM is long because it usually involves many science frames (it can take up to a few tens of seconds to minutes for a single FM depending on the number of wavelengths and frames in the sequence). Therefore, we do not recommend to use this in an MCMC wrapper exploring all parameters at all wavelengths. If one wants to do disk spectroscopy using this method we recommend:
\begin{enumerate}
\item Stack each IFS cubes in a single larger bandwidth frame.
\item Fit the best model, treating the stacked frames as a single-wavelength sequence.
\item Duplicate your best model to create a multi-wavelength cube.
\item Use this model in {\tt DiskFM} to create a multi-wavelength FM corresponding to the initial IFS cubes sequence. 
\item Finally only adjust the photometry in each of the IFS sub-wavelength.
\end{enumerate}
This method assumes that in within an IFS bandwidth, all parameters (disks geometry and SPFs) remain constant and that only the photometry varies. This also relies on the assumption that the FM is a linear function of the total photometry. This linearity assumption, also used for planets\cite{pueyo_detection_2016}, has been verified for disks with {\tt DiskFM}. We did not show any {\tt DiskFM} performance tests on disk spectroscopy in this paper.

\section{Performance of {\tt DiskFM}}
\label{sec:performance}
In this section we show performance of this code. We used a common approach to test and show performance in planet detection algorithms: inject a disk in an empty sequence and try to recover its parameter with the smallest error bars. With this method tested {\tt DiskDM} in 30 different cases: different geometries (3), different photometries (one ``bright" and one ``faint"), different observation modes (ADI and RDI) and finally different reduction parameters (conservative to aggressive).

\subsection{Description of the test method}

We chose 3 well known disks with different geometries to illustrate the performance of the code (see our models Fig.~\ref{fig:show_oversub_selfsub}, top line): 
\begin{itemize}
\item the almost face-on HD 181327\cite{schneider_discovery_2006} (i $\sim$ 30\textdegree), currently undetected from the ground, probably partly due of ADI self-subtraction.
\item the almost edge-on HD 32297\cite{Schneider2005} (i $\sim$ 88\textdegree), with a very high signal-to-noise ratio, which makes it very sensitive to self-subtraction
\item the iconic HR 4796 A\cite{schneider_stis_2008} (i $\sim$ 77\textdegree), with a unique projected radius and inclination which allow the precise analysis of the SPF, both in polarized and total intensity. 
\end{itemize}

\subsubsection{Create a realistic data set}

To create realistic sets of data, we used a long GPI H-Spec observational sequence with 56 observations over an hour (60 seconds each) with a large range of parallactic angles ($\Delta$PA = 85.4\textdegree). The sequence was obtained with GPI, using the star HD 48525, on January 28\textsuperscript{th}, 2018, and does not contain any astrophysical signal to the best of our knowledge (hereafter ``empty sequence"). We cut this sequence in 2 (28 frames in each group, 60 seconds each). The first sequence was used as the ``science observation" sequence (28 frames, $\Delta$PA = 59.4\textdegree) in which we injected the modeled disks, the second was used as a reference sequence (28 frames). This allowed us to simulate both ADI and RDI reductions for each of our modeled disks. 

Each GPI-Spec observation includes four satellite spots \cite{wang_gemini_2014}. We approximated the PSF during each observation by averaging the images of the four satellite spots to increase the SNR. This PSF was used to convolve all of the modelled disks before injecting them in the sequence.

For RDI, we use 3 KL modes from conservative to aggressive (KL\# 5, KL\# 10, KL\# 20). For ADI we also used 3 KL modes from conservative to aggressive (KL\#5, KL\# 10, KL\# 20), except for the ADI reduction of the face-on disk HD 181327 where we tried to stay as conservative as possible (KL\# 1, KL\# 3, KL\# 5) in order to detect the disk in spite of self-subtraction. The exclusion angle parameter (minimum angle between the disk in the science frame and the disks in the PSF library) was set to 6\textdegree. 
We did a full frame reduction and did not introduce smaller reduction zones in this work. Our goal for this test is to show that {\tt DiskFM} associated in an MCMC wrapper can accurately recover the injected parameters to within $1 \sigma$ of the values assumed.

\begin{figure}
\begin{center}
 \includegraphics[width = 0.5\textwidth]{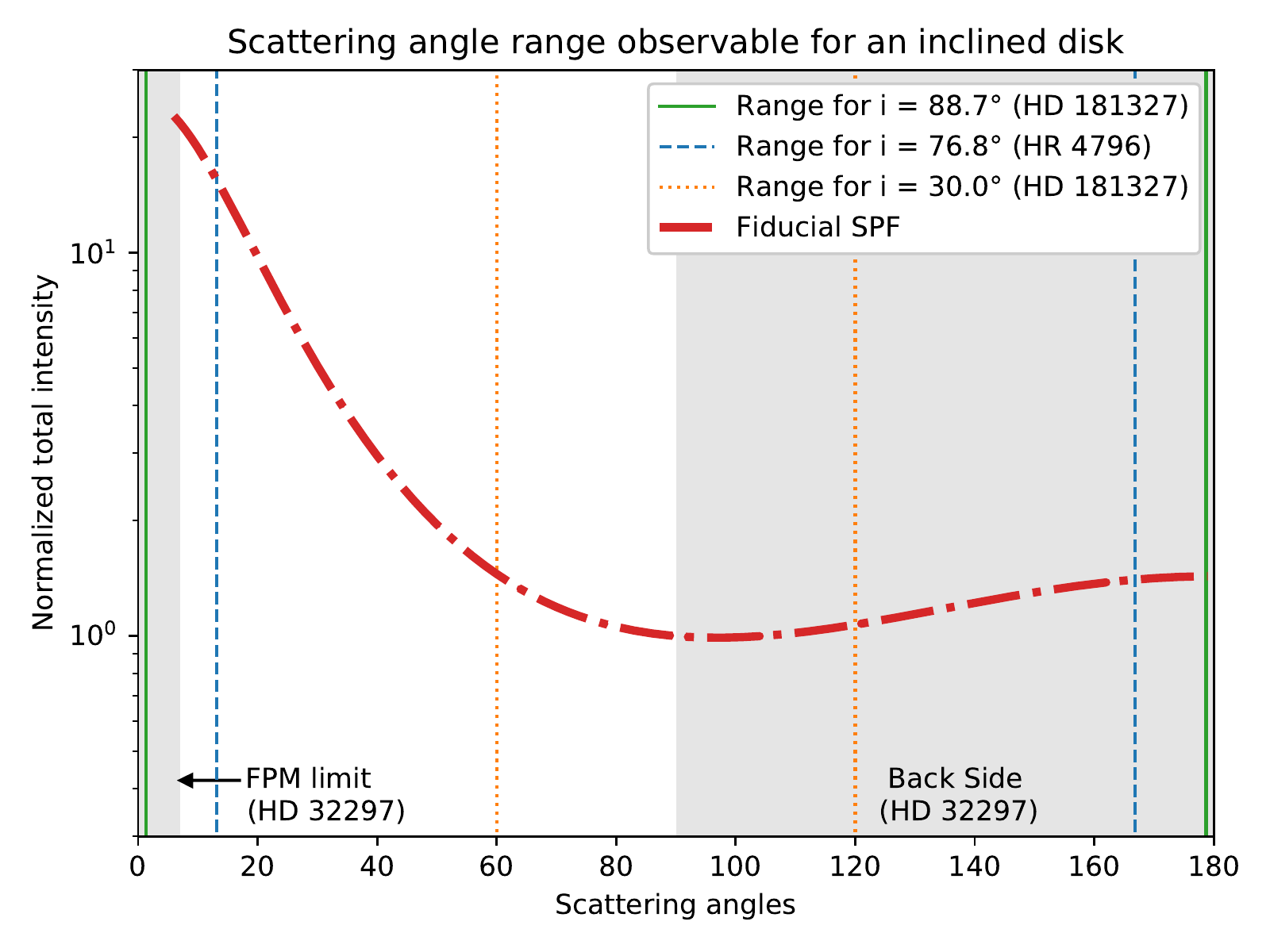}
 \end{center}
\caption[plop]
{\label{fig:SPF_recovered_for_all_geometries} Range of scattering angles that we can probe, depending on the inclination of the observed disk. Scattering information outside of this range cannot be accessed. For some disks (HD 32297 here), other constraints can limit the SPF retrieval, such as parts of the disk that are behind the focal plane mask (FPM) or on the back side of the disk, which is masked by the front side of the disk for this high inclination.}
\end{figure}

\subsubsection{Description of the injected model disks}

For these 3 disks we simulate a realistic model using the method described in Millar-Blanchaer et al. (2016) \cite{millar-blanchaer_imaging_2016}
This code, used both to model our injected disks and to recover them, is a three-dimensional dust density model described by a radial power law, a Gaussian height profile, and a constant aspect ratio (that we set here at 1\%). Optically thin scattering is assumed and a 2 component Henyey–Greenstein (HG) function is used as the SPF:
\begin{equation}
\label{eq:2g_spf}
SPF_{g1,g2,\alpha}(\theta) = \alpha HG_{g1}(\theta) + (1-\alpha) HG_{g2}(\theta)
\end{equation}
where $HG_{g}$ is the one component HG function of parameter g. We use zodiacal dust parameters measured by Hong (1985)\cite{hong_henyey-greenstein_1985}: g1 = 0.7, g2 = -0.2 and the weighting parameter $\alpha$ = 0.66. This SPF is plotted in Fig.~\ref{fig:SPF_recovered_for_all_geometries} (red dash-dotted lines) and is representative of the SPF currently observed for most debris disks \cite{hughes_debris_2018}. This SPF is normalized at 90\textdegree (disk ansae), and we multiply the disk by an ``absolute photometry" parameter $N$. We use an inner ($R1$) and outer radius of the disk ($R2$). The dust density decreases with a slope $\beta_{out}$ between these 2 radii. Inside $R1$, the dust density decreases with a slope $\beta_{in}$. We introduce stellar positional offsets, $dx$ (along the minor axis measured in au in the disk plane) $dy$ (along the major axis measured in au in the disk plane). Finally, we complete our model by including two observation parameters, inclination $i$ and position angle $PA$. 
We used reasonable parameter values found in the recent literature for each disk \cite{stark_revealing_2014,duchene_gemini_2020,chen_multiband_2020}. The value sets for these parameters (hereafter injected disk ``True" parameters) for each disks can be found in the bottom lines of Tables \ref{tab:resultats_mcmc_hd181}, \ref{tab:resultats_mcmc_hd32} and \ref{tab:resultats_mcmc_hr47}. The resulting models (after convolution by the PSF os the instrument) can be found in Fig.~\ref{fig:show_oversub_selfsub} (top line).

For the face-on disk HD 181327, we had little hope to recover the disk in the ADI-reduced data and the analysis was mainly focused on RDI reduction analysis. Because over-subtraction (the only bias in RDI) does not depend on the disk signal, we only introduced the disk at a single absolute photometry: hereafter HD 181327-like disk. Additionally, because of the low inclination, we did not expect to extract much information from the SPF: the range of scattering angles that we can probe depends on the inclination of the observed disk and is very small for i = 30\textdegree (see orange dotted vertical lines in Fig.~\ref{fig:SPF_recovered_for_all_geometries}). Therefore, we fixed the SPF parameters and only attempted to recover 9 parameters ($R1$, $R2$, $\beta_{in}$, $\beta_{out}$, $i$, $PA$, $dx$, $dy$ and $N$).

For the 2 other disks (HR 4796 A and HR 32297), we specifically wanted to test the effects of absolute photometry on self-subtraction, calculating the disk models at two different photometries, producing 4 different models: hereafter, ``bright" HR 4796 A-like disk, ``faint" HR 4796 A-like disk, ``bright" HD 32297-like disk, ``faint" HD 32297-like disk. For these disks we left 8 geometric parameters as free values ($R1$, $R2$, $\beta_{out}$, $i$, $PA$, $dx$, $dy$, $N$), as well as 3 free parameters for the SPF ($g1$, $g2$, $\alpha$), for a total of 11 free parameters. $\beta_{in}$ was fixed because this was considered a difficult parameter to fit for high inclination disks. 

\begin{figure}
\centering  \includegraphics[width=0.8\textwidth, trim=0 80 0 0, clip]{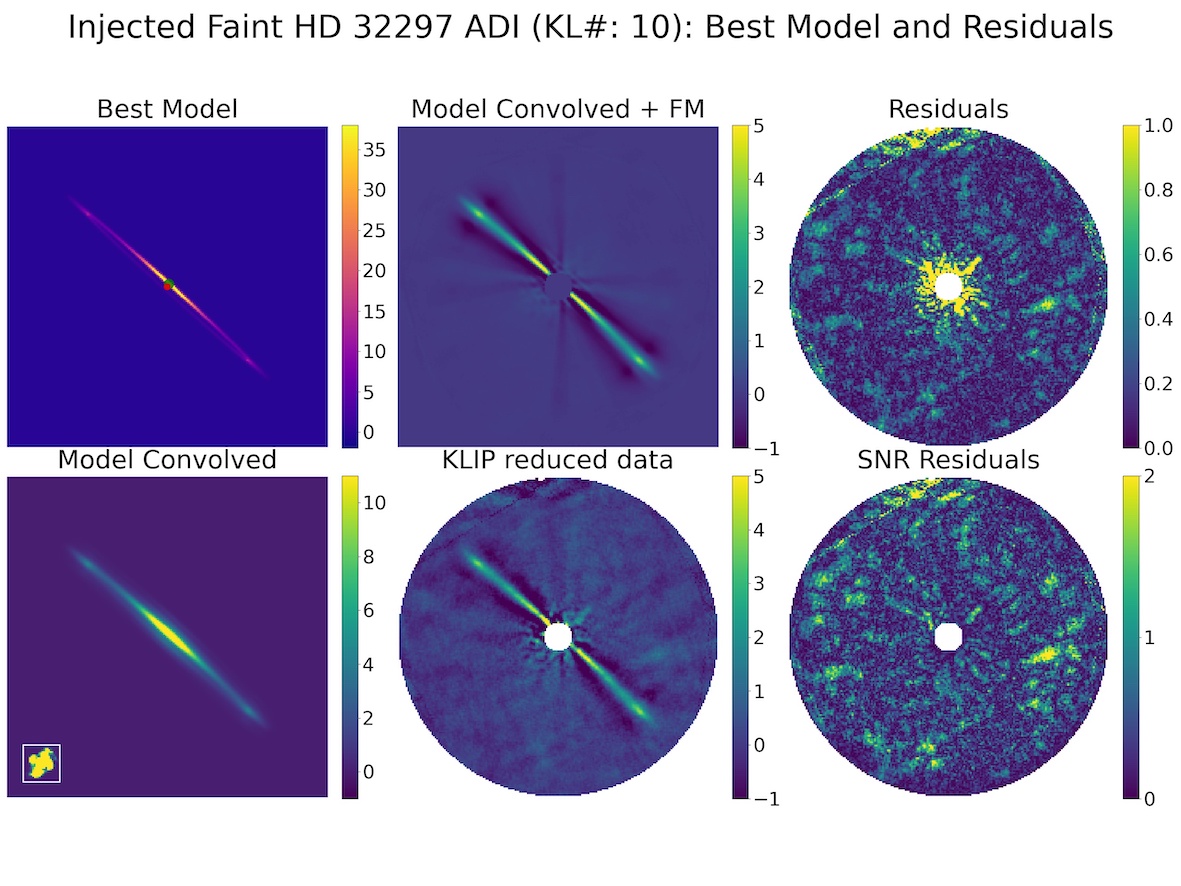}
\caption[plop]{
 \label{fig:Fake_hd32297faint_adikl10_backend_file_mcmc_BestModel_Plot}  Best-fit model resulting from the MCMC for the ``faint" HD 32297-like disk with ADI reduction and a KL\# 10. (Top-Left) Best-fit model image; the red and green spots mark the position of the star and the disk pericenter, respectively. (Bottom-Left) Best-fit model image after convolution with an observed GPI PSF (shown in a small vignette at the Bottom-Left). (Top-Middle) Best-fit model image after convolution and FM to reproduce KLIP effect. (Bottom-Middle) Image showing the KLIP-ADI reduced dataset. (Top-Left) Residuals from the MCMC. (Bottom-Right) SNR of the residuals of the MCMC.}
\end{figure} 

In total, we produced 5 disks * 2 observation modes (ADI, RDI) * 3 parameter sets (3 different KL numbers) = 30 sequences reduced by KLIP to analyze with {\tt DiskFM} in a Bayesian analysis to extract parameters. 

\subsubsection{Bayesian parameter estimation}

\begin{figure}
\begin{center}
 \includegraphics[width = \textwidth]{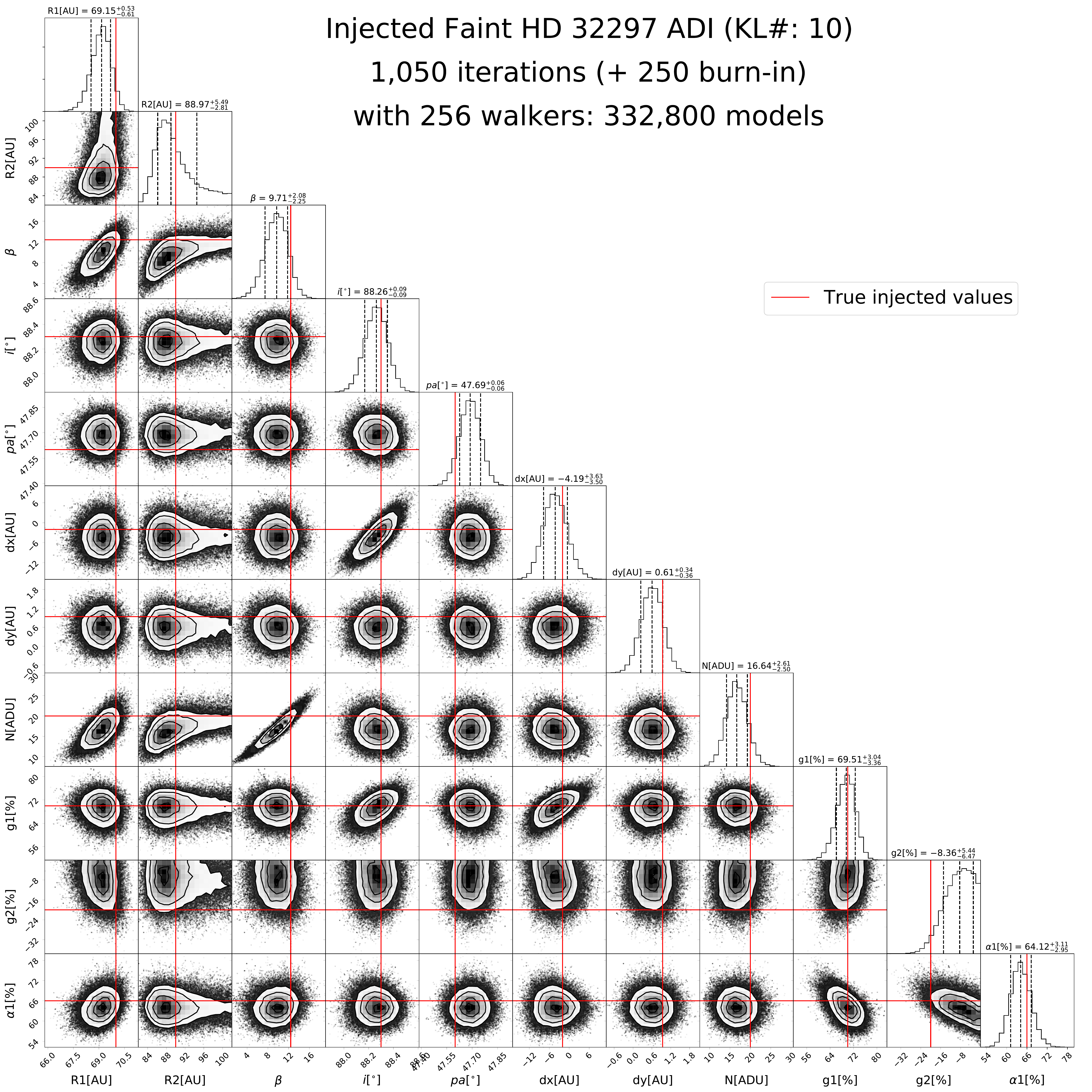}
 \end{center}
\caption[plop]
{\label{fig:Fake_hd32297faint_adikl10_backend_file_mcmc_pdfs} MCMC posterior distributions recovered for the ``faint" HD 32297-like disk with ADI reduction and 10 KL modes used. The diagonal histograms show the posterior distributions of each parameter marginalized over all other parameters. In each plot, the dashed lines show the 16\textsuperscript{th}, 50\textsuperscript{th}, and 84\textsuperscript{th} percentiles. The off-diagonal plots display the joint probability distributions with contour levels at the same percentiles. For each posterior, the parameter “true” value of the injected disk is over-plotted in red lines.}
\end{figure}

The detailed process is described in Chen et al. (2020)\cite{chen_multiband_2020} and we will quickly recall the main features. For each sequence, we used Bayesian parameter estimation to derive the best fit values and the Posterior Distribution Functions (PDFs) of each parameters. We used the following steps:

\begin{enumerate}
\item We randomly pick a set of free parameters.
\item We generate a model image of the disk (Fig.~\ref{fig:Fake_hd32297faint_adikl10_backend_file_mcmc_BestModel_Plot}, Top-Left)
\item We convolve the model with the PSF (Fig.~\ref{fig:Fake_hd32297faint_adikl10_backend_file_mcmc_BestModel_Plot}, Bottom-Left)
\item We FM the convolved model using {\tt DiskFM} to simulate the impact of the ADI/KLIP or RDI/KLIP on the model (Fig.~\ref{fig:Fake_hd32297faint_adikl10_backend_file_mcmc_BestModel_Plot}, Top-Middle). 
\item We measure a likelihood by comparing the FM to the reduced image (shown in Fig.~\ref{fig:Fake_hd32297faint_adikl10_backend_file_mcmc_BestModel_Plot}, Bottom-Middle) by measuring:
\begin{equation}
\chi^2 =\sum \frac{(Data - ForwardModel)^2}{Uncertainty^2}
\end{equation} 
The uncertainty is measured with a technique described in Chen et al. (2020)\cite{chen_multiband_2020}.
\end{enumerate}

Note that for step 3, the PSF is measured directly from the satellite spots, in GPI-Spec observations. However, the disk rotated onto the satellite spots in some exposures. As we would do in an analysis on real GPI data, we omitted the affected satellite spots from the PSF estimate to keep the PSF as accurate as possible. The resulting PSF is shown in a small vignette on the Bottom-Left of Fig.~\ref{fig:Fake_hd32297faint_adikl10_backend_file_mcmc_BestModel_Plot}. For this reason, the PSF used in Step 3 above is slightly different from the one actually used to convolve the injected model. This was done purposefully to introduce a small source of error to the PSF and make it more realistic.

We performed these steps hundreds of thousands of times within an MCMC wrapper that maximizes $e^{-\chi^2/2}$ until the chains converged, using the {\tt emcee} package \cite{foreman-mackey_emcee:_2013}. We used 256 parallel walkers and removed some iterations during the burn-in phase. For this test, we used the ``True" parameter value as initial points to reduced the burn-in phase. For each case, we ran 250 iterations for burn-in and then at least 1000 iterations. The priors were chosen flat and relatively large, centered around the ``True" parameter values. Finally, we plot the residuals for the best model (Fig.~\ref{fig:Fake_hd32297faint_adikl10_backend_file_mcmc_BestModel_Plot}, Top-Right), the SNR of the residuals (Fig.~\ref{fig:Fake_hd32297faint_adikl10_backend_file_mcmc_BestModel_Plot}, Bottom-Right), as well as the Posterior Distribution Functions (Fig.~\ref{fig:Fake_hd32297faint_adikl10_backend_file_mcmc_pdfs}). We derived uncertainties based on the 16\textsuperscript{th} ($-1\sigma$), 50\textsuperscript{th} (median value), and 84\textsuperscript{th} ($+1\sigma$) percentiles of the samples in the distributions (plotted as vertical lines in the corner plots). For each posterior, the parameter “True” value of the injected disk is over-plotted in red lines. We measure the performance of {\tt DiskFM} for this sequence by its ability to recover the initial True value within $1\sigma$. 

\begin{figure}
\begin{center}
 \includegraphics[width = 0.5\textwidth]{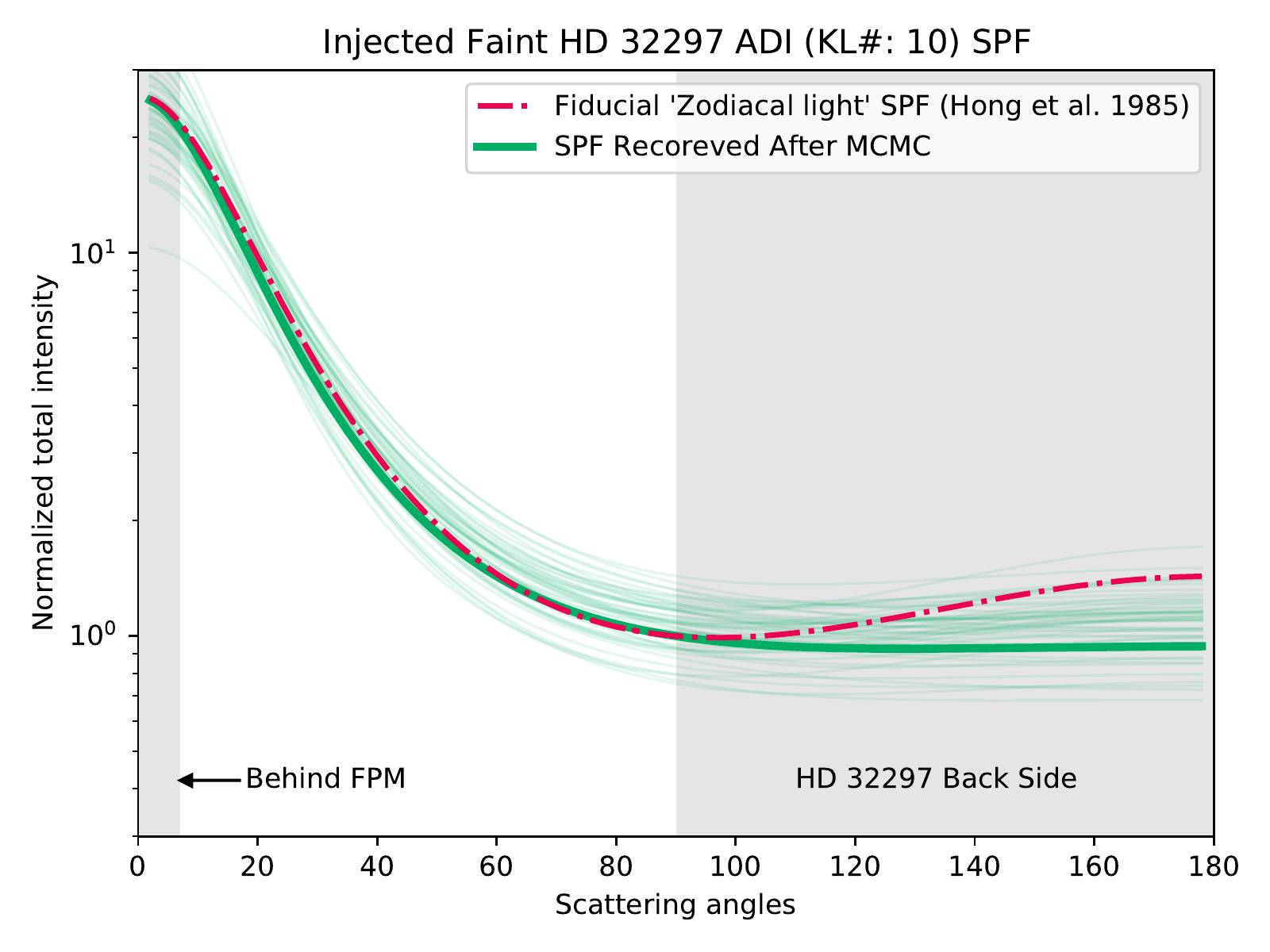}
 \end{center}
\caption[plop]
{\label{fig:Fake_hd32297faint_adikl10_comparison_spf} SPF injected into disk image (red dashed line) compared with the estimated phase function recovered from disk fitting (green solid line) for the ``faint" HD 32297-like disk with ADI reduction and a 10 KL modes used. We draw 50 random SPFs from the MCMC sampler, after convergence of the MCMC, that we plot in green to estimate uncertainty.}
\end{figure}

Finally, in Fig.~\ref{fig:Fake_hd32297faint_adikl10_comparison_spf}, we show the injected SPF (red dashed line) and recovered SPF (green solid line). We draw 50 random SPFs from the MCMC sampler, after convergence of the MCMC, that we plot in green to estimate uncertainty. 

For each of the 30 cases, we produced at least 256 walkers * (1250 iterations) = 320,000 (models + FM), which took less than 24h using 48 CPUs on one of Paris Observatory clusters (for a total of 10 million models + FM which took around 10 days).

\subsection{Results}

\begin{table}
\begin{center}
\vspace{0.5cm}
 \includegraphics[width = \textwidth]{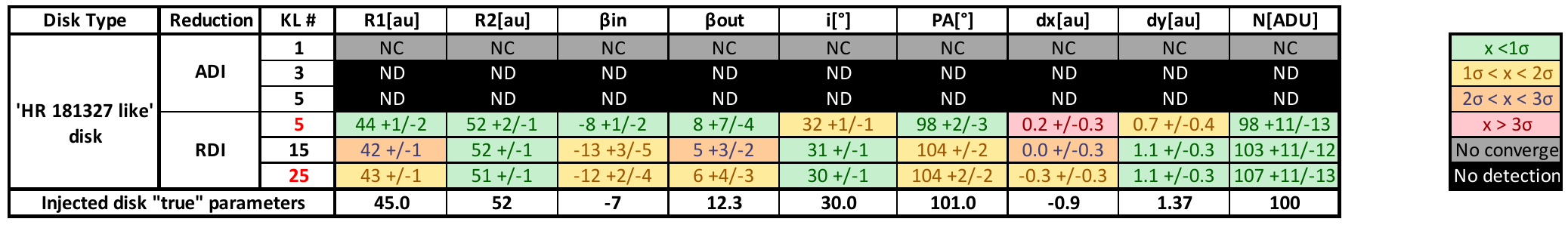}
 \vspace{-0.8cm}
 \end{center}
\caption[plop]
{\label{tab:resultats_mcmc_hd181} Results for the HD 181327-like disk. The best cases are RDI with 25 KL modes used (more aggressive).}
\end{table}

In the last section, we deliberately showed one of the case where {\tt DiskFM} was not the most accurate: ADI reduction of the Faint HD 32297 like disk with KL 10 (slightly aggressive reduction). Fig.~\ref{fig:Fake_hd32297faint_adikl10_backend_file_mcmc_BestModel_Plot} shows that the disk is well recovered, the residuals do not show any residuals of the disk. Additionally, Fig.~\ref{fig:Fake_hd32297faint_adikl10_backend_file_mcmc_pdfs} shows that the MCMC has converged normally with Gaussian posteriors and that most parameters are recovered  within $1\sigma$. However, 5 parameters ($R1$, $\beta_{out}$,  $PA$, $N$ and $g2$ have not been recovered). For some of these parameters, we can probably blame the geometry of the disk: the parameter $g2$ is describing the SPF in the back side of the disk. Because of the very high inclination of this disk, the front and back sides of the disk are closer than the resolution of the telescope. As a result, most of the information on the back side of the disk is lost during observation (see disk convolved by the telescope PSF in Fig.~\ref{fig:Fake_hd32297faint_adikl10_backend_file_mcmc_BestModel_Plot}, bottom left). Fig.~\ref{fig:Fake_hd32297faint_adikl10_backend_file_mcmc_pdfs}) shows that the SPF of the disk has been well retrieved in the front side of the disk. Other parameters, like the inner radius R1, should have been recovered despite the challenging geometry: the recovered value found $R1 = 69.15 +0.53 /-0.61$ au, therefore the ``True" value (70.0 au) is 1.6 $\sigma$ away from the recovered value. This shows that {\tt DiskFM} is producing slightly narrow error bars for this geometry, brightness, reduction method and set of reduction parameters. In Annex \ref{sec:best_and_worst}, we show some of the best and worst results of our tests. 

In this section we gathered the results in Tables 1 to 3. Each cell of these tables shows the recovered value with $1 \sigma$ error bars. The true values are recalled at the bottom of each table. 

The disk was detected in all but two instances, for the almost face-on HD 181327-like disk in ADI, with KL numbers equal to 3 and 5 (black lines in Table \ref{tab:resultats_mcmc_hd181}). This is not surprising, as self-subtraction is an expected effect of ADI. The MCMC returned random flat posteriors. In two other instances, the face-on HD 181327-like disk in ADI (with KL\# 1) and the ``faint" HD 32297 like disk RDI (with KL\# 5), the disk was barely detected by visual inspection but not constrained by the MCMC (flat posterior or posterior stacked against one of the prior limit). For the face-on HD 181327-like disk with ADI KL\# 1, this is for the same reason as previously described (self-subtraction heavily impacts face-one disks). This detection is shown in Fig.~\ref{fig:best_worst_hd181327} in the Appendix. For the ``faint" HD 32297-like disk with RDI KL\# 5, we can assume that the KLIP/RDI with conservative reduction parameters does not remove enough speckles to analyze the disk properly. We note that for more aggressive reduction parameters, at least some parameters of the disk are well constrained.

\begin{table}
\begin{center}
    \vspace{0.5cm}
 \includegraphics[width = \textwidth]{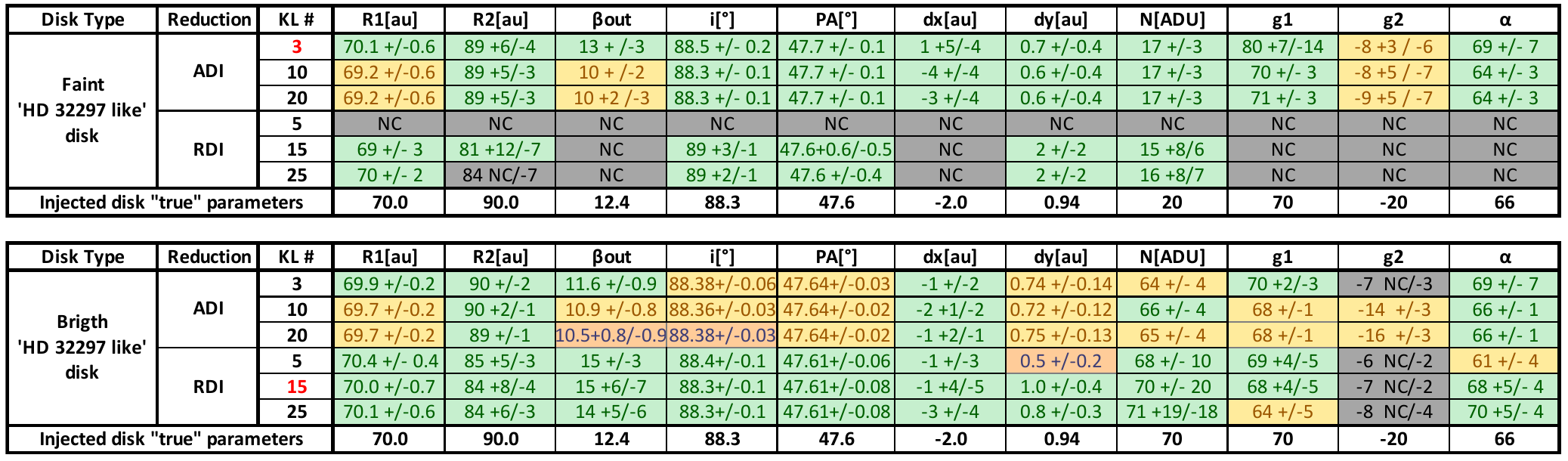}
 \vspace{-0.8cm}
 \end{center}
\caption[plop]
{\label{tab:resultats_mcmc_hd32} Results for the HD 32297-like disks. The best cases are ADI KL\# 3 (conservative) for the faint disk and RDI KL\# 15 (relatively aggressive) for the bright disk.}
\end{table}

Apart from these 4 sequences, we categorized the estimation of the other parameters in 4 categories: good fit (``True" value within $1 \sigma$, green), medium fit (``True" value within $2 \sigma$, yellow), bad fit (``True" value within $3 \sigma$, orange), dreadful situation (``True" value larger than $3 \sigma$, red) and parameter unconstrained (grey). This last category corresponds to flat posteriors or posteriors constrained in only one direction. In practice, finding an unconstrained parameter would warrant re-running the MCMC using a simpler model where this parameter is fixed or constrained with a more narrow prior. Note that unconstrained results are a lot less worrisome than well-converged parameters with estimated values different from ``True" value by more than $3 \sigma$ results (red) which would lead to publications of errors on these parameters. This is fortunately very rare. 

In practice, these kinds of ``inject and recover" tests should be done in most studies to test, for a given disk, what is the best course of action, as this process is done for point sources in planet studies. We can only give broad guidelines that will help {\tt DiskFM} users to make first guesses at the method or aggressiveness to analyze their own data:
\begin{itemize}
\item ADI is not well suited for face-on disks, as expected.
\item Some parameters are extremely difficult to recover, like one of the offsets in the HD 181327-like disk or the $g2$ parameter in the HD 32297-like disk.  
For these kinds of parameters, we can speculate if we could recover them in an ideal post-processing case or if this information is already unrecoverable because of telescope resolution. A good test would be to try to extract these parameters directly from a convolved model.
\item For ADI, the best reduction is almost always the most conservative. In all cases but one, the best reduction is the most conservative reduction (low KL number) parameters.
\item ADI gives better results on faint disks than on bright disks. We suspect that self-subtraction more heavily impacts bright disks. For these bright disks, RDI could be a better option in some cases.
\item {\tt DiskFM} currently gives under-estimated error bars in some cases. This could be fixed by slightly increasing the noise levels (the method we currently used assumes speckle noise distribution to be Gaussian in the uncertainty map, which is not accurate.) Note that a multiplication by 2 of all error bars (yellow cells would become green, orange and most reds would become yellow) would lead to much improved results of {\tt DiskFM}. The estimation of a correct uncertainty map in coronagraphic images is not a problem specific to this algorithm nor to disk imaging\cite{pairet_stim_2019}).
\end{itemize}

\begin{table}
\begin{center}
\vspace{0.5cm}
 \includegraphics[width = \textwidth]{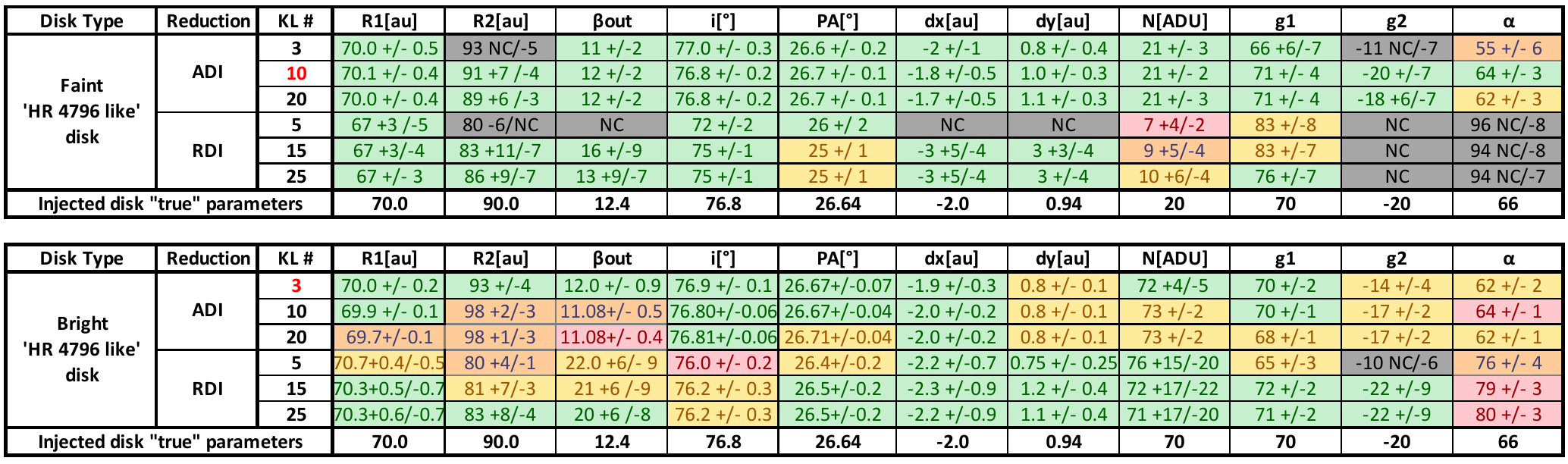}
 \vspace{-0.8cm}
 \end{center}
\caption[plop]
{\label{tab:resultats_mcmc_hr47} Results for the HR 4796 A-like disk. The best cases are ADI KL\# 10 (relatively aggressive) for the faint disk and ADI KL\# 3 (conservative) for the bright disk.}
\end{table}

\subsection{Limits of this test study}

In this section, we identify a few limits of this study. First, the initial setup of the RDI sequences could have biased these results. Because the reference library used for RDI uses images from the same observing sequence as the science sequence, the reconstructed reference PSF may have been too perfect. In practice, for Ground-based observations, References library are assemble using PSFs from different stars over a larger period of time. Actual RDI sequences with Ground-based observations might be much more complicated by the fact that the PSF usually varies a lot depending on observing conditions. 

From our own experience\cite{chen_multiband_2020}, we know that the quality of the PSF can have a huge impact on the quality of the fit. We tried to artificially slightly degrade our PSF, but in real cases, the difference between the measured PSF and the PSF used while modeling might be larger, leading to larger error bars. Once again, this is not a problem specific to {\tt DiskFM} nor to disk imaging in general.

In these simple tests, the model science images were generated using the same modelling code that was used to simulate the disk during the MCMC retrievals, which ensured that the disk in the data could be modeled (the injected disk was within reach in our search). This is probably not the case in real data, where the disks are always more complicated than our simple models. 

\section{Conclusion}

In conclusion, {\tt DiskFM} is a powerful tool to extract physical parameters from a coronagraphic image of a disk in ADI, RDI or SDI. This proceeding shows that {\tt DiskFM} can sometimes gives under-estimated error bars, which encourage us to estimate more carefully uncertainty maps. We encourage developers of other disk post-processing techniques to reproduce this analysis with their own methods. The planet imaging community is now organizing detection and characterization challenges\footnote{e.g. the Exoplanet Imaging Data Challenge: \url{https://exoplanet-imaging-challenge.github.io/}} that could be imported in the circumstellar disk community, following the method that is shown in this paper, with the goal of aligning our metrics, and goals.

\appendix    

\section{Best and Worst results of {\tt DiskFM}}
\label{sec:best_and_worst}
\begin{figure}
\begin{center}
 \includegraphics[width =0.8\textwidth]{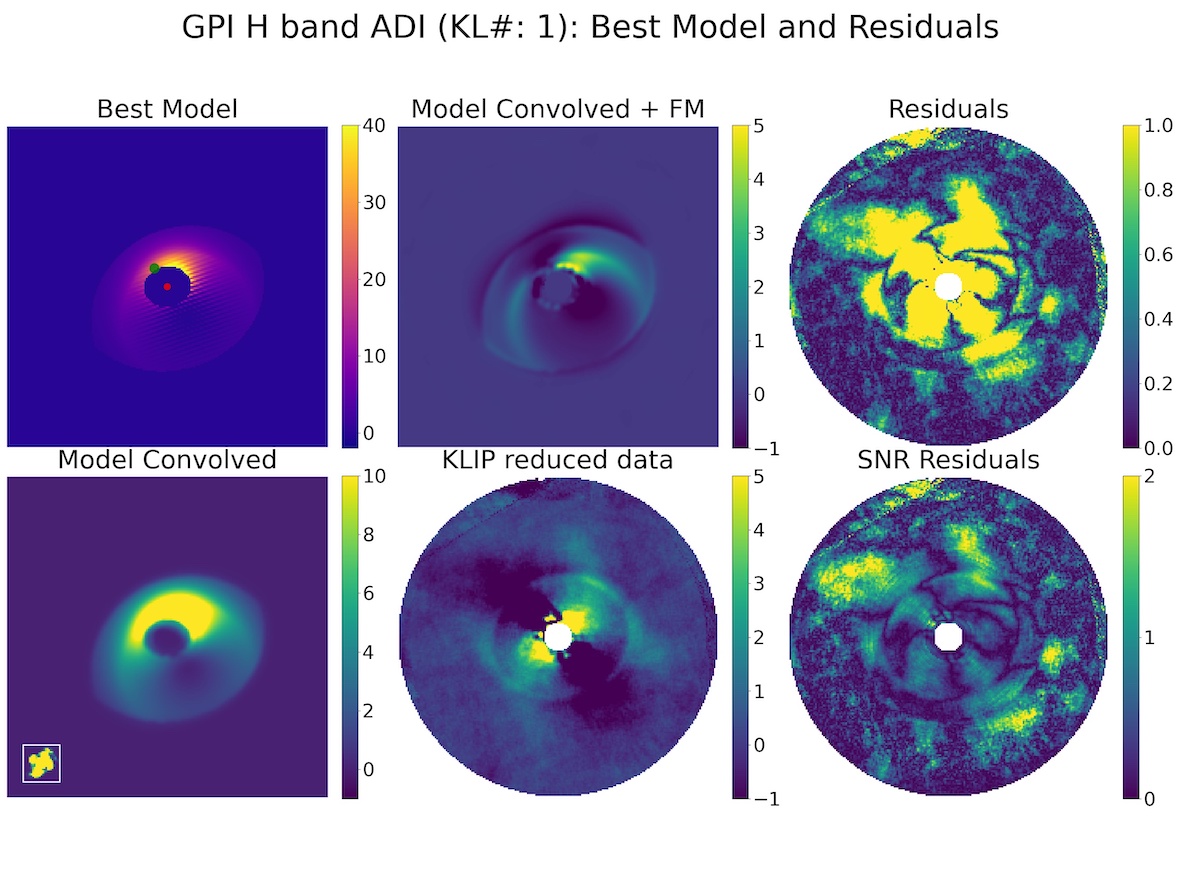}
 \includegraphics[width = 0.8\textwidth]{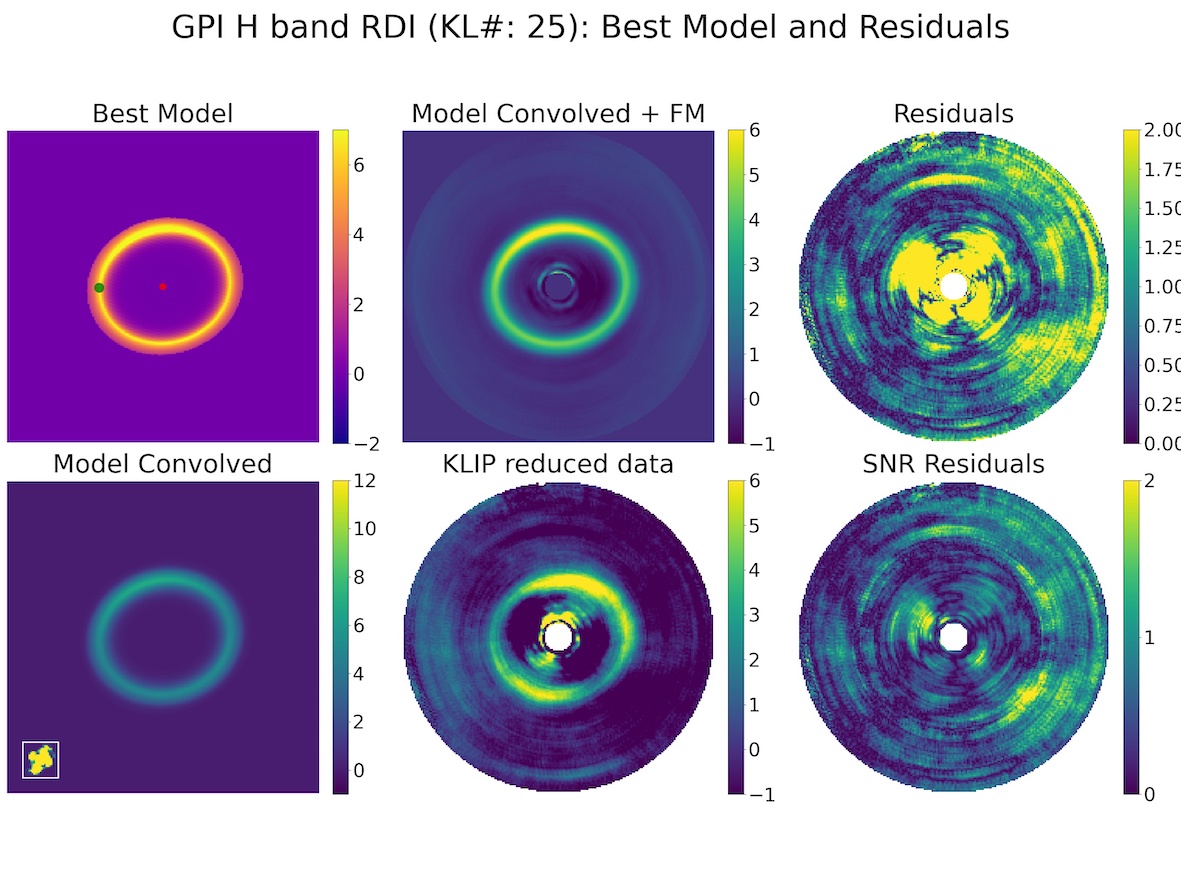}
 \end{center}
\caption[plop]
{\label{fig:best_worst_hd181327} Best and worst cases for HD 181327. \textbf{Top:} The disk is barely detected in ADI, even for the most conservative parameters (KL\# 1). The best fit does not reproduce the data at all. \textbf{Bottom:} In Aggressive RDI, the disk is mostly extracted.}
\end{figure}

\begin{figure}

 \centering \includegraphics[width =0.77\textwidth]{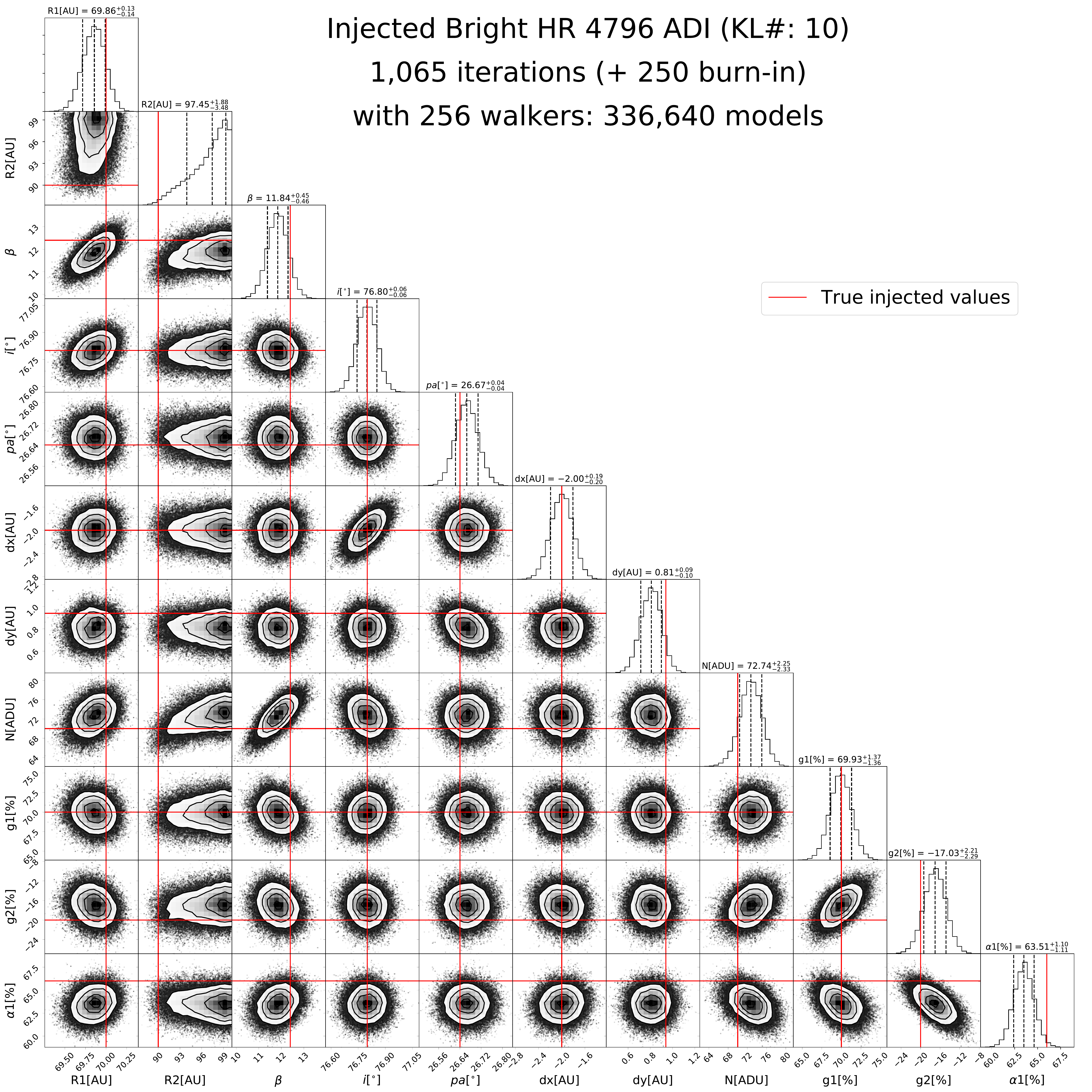}\color{white}{.}
 
 \vspace{-11.5cm}\color{white}{.}\hspace{8.5cm}\color{white}{.}
 \includegraphics[width =0.38\textwidth]{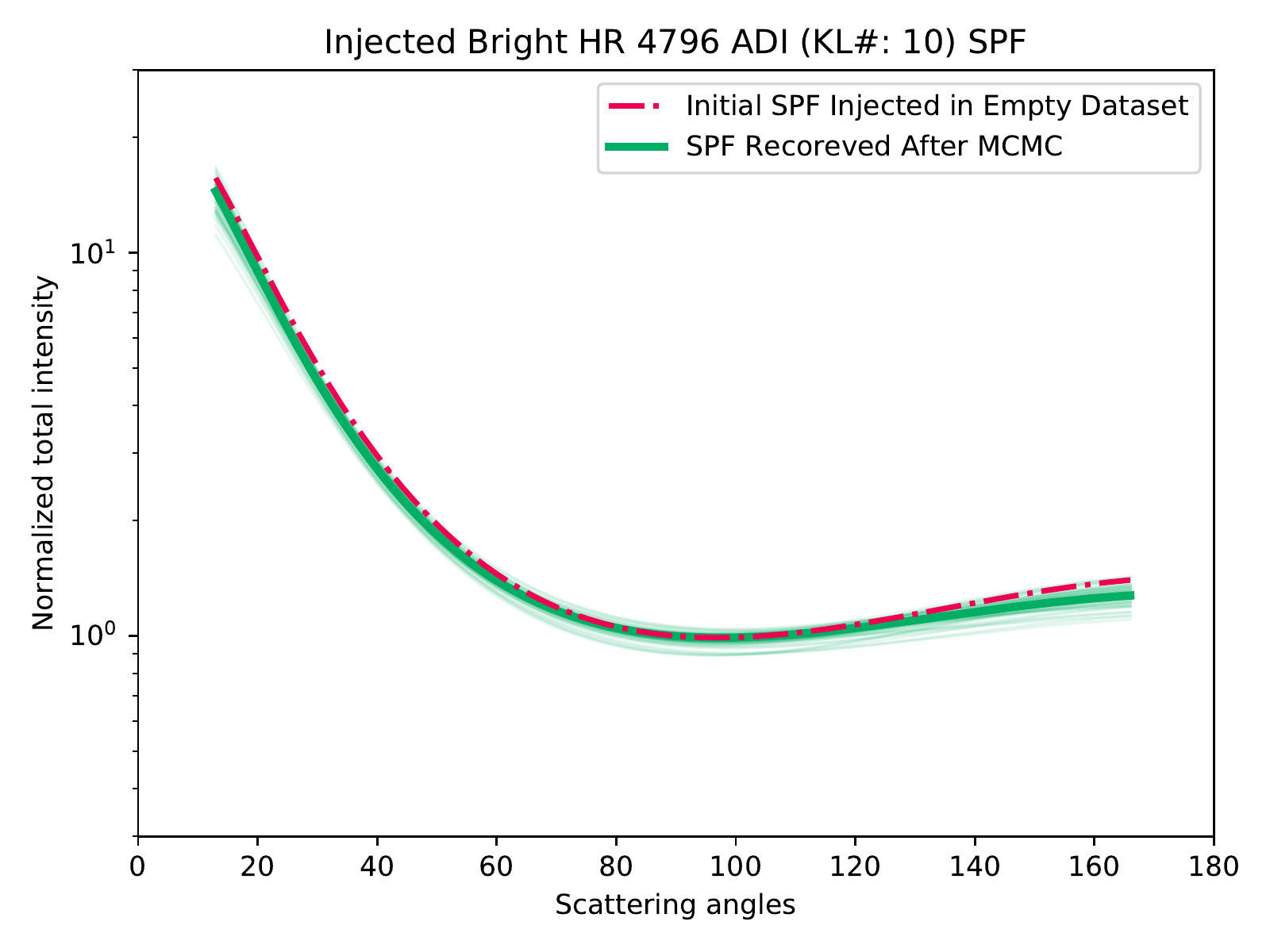}
 \color{white}{.}\vspace{6.6cm}\color{white}{.}
 \includegraphics[width =0.7\textwidth]{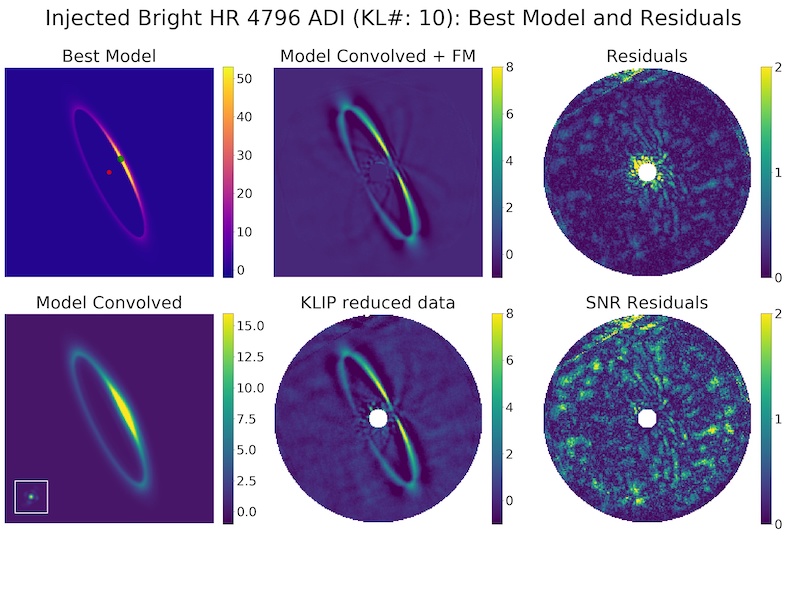}
  \vspace{-0.9cm}
\color{black}{\caption[plop]
{\label{fig:bright_hr4796_adi_kl10} Bright HR 4796 A-like disk, ADI reduction (KL\# 10). SPF is recovered but important self-subtraction effects leave noticeable residuals.}}
\end{figure}

This appendix shows some interesting {\tt DiskFM} MCMC results. Fig.~\ref{fig:best_worst_hd181327} shows results for HD 181327 in ADI and RDI. Fig \ref{fig:bright_hr4796_adi_kl10} shows the difficult case of the bright HR 4796 A disk.

\acknowledgments 
 
 This material is based upon work supported by the National Science Foundation under Astronomy and Astrophysics Grant No.~1616097 (JM). JM acknowledges support for part of this work was provided by NASA through the NASA Hubble Fellowship grant \#\textit{HST}-HF2-51414 awarded by the Space Telescope Science Institute, which is operated by the Association of Universities for Research in Astronomy, Inc., for NASA, under contract NAS5-26555. This work uses observations obtained at the Gemini Observatory.

Software: Gemini Planet Imager DRP\cite{perrin14, perrin16}, {\tt pyKLIP}\cite{wang_pyklip_2015} , {\tt numpy, scipy, Astropy} \cite{astropy2018}, {\tt matplotlib} \cite{matplotlib2007, matplotlib_v2.0.2}, {\tt emcee}\cite{foreman-mackey_emcee:_2013}, {\tt corner}\cite{corner18}

\bibliography{biblio_spie} 
\bibliographystyle{spiebib} 

\end{document}